\documentclass{JAC2000}
\usepackage{graphicx}
\begin{document}
\title{New 357 MHz Sub Harmonic Buncher}
\author{M. Kuriki, H. Hayano, K. Hasegawa$\rm ^{a}$\\
High energy accelerator research organization, Tsukuba, Ibaraki, Japan\\
(a)Scientific university of Tokyo, Noda, Chiba, Japan}
\maketitle

\begin{abstract}
KEK-ATF is doing R\&D establishing high-current and low-emittance
electron beam for the future linear colliders. In ATF, 15 ps length
electron beam is generated by combination of a thermionic gun, a
couple of 357 MHz SHBs, and a S-band TW buncher. In the linac, beam
instability caused by the fluctuated amplitude and phase of SHBs has
been observed.  The reason of the fluctuation is considered to be the
multi-pacting in the SHB cavity. In addition, Q value of the cavity is
almost half of the designed value, then SHBs are not operated in an
optimum condition.  We have fabricated new SHBs with considerations
avoiding the multi-pacting and recovering Q value. A simulation shows
that bunching quality is improved with the new SHB cavity.

\end{abstract}

\section{Introduction}
\vspace{-.2cm}
\label{sec:one} 
KEK-ATF is a test facility to study the low-emittance multi-bunch beam
and beam instrumentation technique for the future linear collider.
That consists from 1.5 GeV S-band linac, a beam transport line, a
damping ring, and a diagnostic extraction line.

In the linac, the electron beam is generated by a thermionic electron
gun. Typical intensity is $\rm 10^{10}$ electron/bunch. The bunch length
shrinks from 1 ns to less than 20 ps by passing a couple of sub-harmonic
bunchers and a TW buncher.  The electron beam is accelerated up to 1.3
GeV by 8 of the S-band regular accelerating sections.

In April 2000, we achieved horizontal emittance $\rm 1.3\times 10^{-9}
rad.m $, vertical emittance$1.7\times 10^{-11} rad.m $ (both for
$2.0\times 10^9 electron/bunch$, single bunch mode )\cite{emittance}
which are almost our target. 

In November 2001, we have started the multi-bunch beam operation. 20 of
bunches separated by 2.8 ns are accelerated by one RF pulse. This
multi-bunch method is one of the key technique in the linear collider.
The commissioning was successfully done. $6.0\times 10^{10}$
electron/train was obtained at the extraction line. We now struggle to
develop many instrumentation devices to measure the multi-bunch profile
clearly.

In ATF, the instability of the injection part of the linac has been an
issue.  From a study, the amplitude and phase fluctuation of SHB
cavities are one of the reason of the instability.

\begin{figure}[htbp]
\vspace{-.2cm}
\begin{minipage}{0.27\textwidth}
\includegraphics*[width=1.8in,height=1.5in]{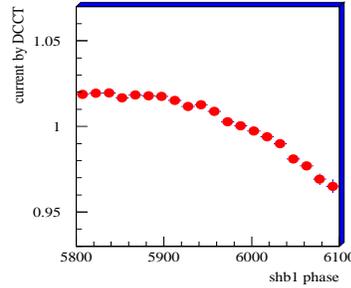}
\end{minipage}
\hfill
\begin{minipage}{0.21\textwidth}
\caption{DR current by DCCT as function of SHB1 phase. Both variables
 are normalized by its average. \label{fig:shb1p}}
\end{minipage}
\vspace{-.4cm}
\end{figure}

Fig. \ref{fig:shb1p} shows the ring current measured by DCCT as function
of phase of SHB1 RF. Both variables are normalized by average of itself. 
From this figure, the SHB1 phase makes the ring current unstable.
Similar property is found on the amplitude and both for SHB2 too. 

In single bunch operation, the injection instability changes only the
beam current. On the ohter hand, in case of multi-bunch operation, it
changes not only the total current, but also the bunch by bunch
profile. Response of most of the instrumentation device such as BPM
depends on the bunch by bunch profile. The fluctuation causes
therefore an extra ambiguity on the measurement. It is currently the
most serious problem of the beam instrumentation for the multi-bunch
beam.

The instability of the bunching system extends the bunch length leading
large energy spread. The injection efficiency to the ring becomes
therefore worse. To examine the stability of the bunching system, the
bunch length is a good index rather than the injection current.

In ATF, a device to measure the bunch length has been placed at the end
of the injector part. This device, so called Bunch Length Monitor, BLM
detects amplitude of 6.5 GHz and 11.4 GHz signals induced by
beam. Bunch length is reconstructed from the both information.

Tab. \ref{tab:stab} shows contributions from various sources on the
bunch length instability. The bunch length instability is evaluated as
the standard deviation normalized by its average.  From these data,
phase of SHB1 and amplitude of SHB2 have large
contributions. Distribution of SHB2 amplitude has two peaks and this
value is calculated for both peaks. If we take only one peak, it becomes
0.0009.
\begin{table}[htbp]
\begin{minipage}{0.15\textwidth}
\caption[]{\footnotesize Contributions from various sources to the bunch length
fluctuation.  The bunch length, $L_b$ is normalized by its average.
\label{tab:stab}}
\end{minipage}
\hfill
\begin{minipage}{0.32\textwidth}
\small
\begin{tabular}{|l|c|}\hline\hline
 Source  & $\displaystyle\frac{\partial L_b}{\partial x}\Delta x$ \\\hline
 Gun HV & 0.0012\\
 SHB1 phase & 0.0067\\
 SHB1 amplitude & 0.0034 \\
 SHB2 phase & 0.0013\\
 SHB2 amplitude & 0.0097 (0.0009)\\\hline\hline
\end{tabular}
\end{minipage}
\vspace{-.5cm}
\end{table}

By monitoring RF in the SHB cavity, RF amplitude distortion is observed
by multi-pacting. This strong multi-pacting can be suppressed by
optimizing magnetic field of Helmholtz coils surrounding the
cavities. Even though, these results suggests that weak multi-pacting
still exists in SHBs and causes the bunch length instability.

\section{New SHB cavity}
\begin{figure}[htbp]
\begin{center}
\includegraphics*[width=3in]{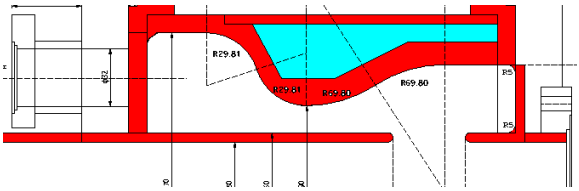}
\vspace{-.2cm}
\caption{Half of cross sectional view of the new SHB cavity. The bottom
of this figure corresponds to the center of the beam
line.\label{fig:shb-cs}}
\end{center}
\vspace{-.3cm}
\end{figure}

\vspace{-.2cm}
To stabilize SHB cavity by avoiding multi-pacting, we have newly made
SHB cavities. Fig. \ref{fig:shb-cs} shows half of the cross sectional
view of the new SHB cavity. The bottom of the figure corresponds to the
center of the beam line. The cavity design was performed with MAFIA
which is a electrical magnetic field simulator.

Comparing to the ordinal pill box cavity, the cavity space is extended
to up-stream direction. This geometry composes $\lambda/2$ structure,
i.e. the left end of the cavity is close (electrically terminated) and
the cavity gap, the right end, works as open end. The benefit of this
geometry was clear if the accelerating gap was placed at the center of
the cavity.  In that case, the cavity length would be twice of the
designed to keep same resonant frequency.

Another tip is the waist at the center of the cavity. Due to the narrow
gap at this waist, the electrical field is concentrated here rather than
on the accelerating gap. It decreases R/Q of the cavity. Because the
beam loading effect is proportional to R/Q, this geometry makes the
cavity strong to the beam loading. It is a big advantage for the
multi-bunch operation. The designed R/Q is 46.2 $\Omega$.

An important issue on the current SHB cavity is low Q value. The
designed Q value is around 8000, but the actual Q value is only
3000. That can be explained by power loss with the contact resistance of
copper gasket. The SHB cavity is composed from three parts, barrel and
end plates. These parts are assembled with copper gasket vacuum
seal. Electrical contact is also made through the gasket with very sharp
edges.

In the new SHB cavity, the copper gasket is still used, but the
electrical contact is made directly on the inner wall. To make a good
electrical contact by keeping vacuum seal, careful turning on the
barrel end where electrical contact is made, was performed. The
designed Q value for the new SHB cavity is 7800, and the measured
value is 7300.  By the careful machining, an ideal electrical contact
was obtained.

\begin{figure}[htbp]
\begin{center}
\includegraphics*[width=3.3in]{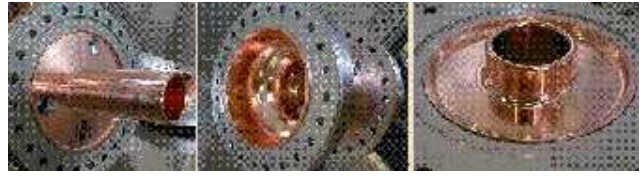}
\vspace{-.7cm}
\caption{Pictures of the new SHB cavity. From left, the up-stream end plate, 
 the barrel, and the down-stream end plate. \label{fig:cavity}}
\end{center}
\vspace{-.4cm}
\end{figure}
Fig. \ref{fig:cavity} shows pictures of the new SHB cavity. From left,
the up-stream end plate which has a long nose corresponding to the beam
pipe, the barrel, and the down-stream end plate. The length of the
cavity along the beam direction is 213 mm. The diameter is varied from
$\phi 170$ to $\phi 90$. The length of the accelerating gap is 40 mm. 

As mentioned in the next section, preventing the multi-pacting is one of
the most important purpose of this new SHB. To suppress it, the cavity
wall is coated by TiN after taking this picture.

\section{Multi-pacting}
\vspace{-.2cm}
Multi-pacting is resonant electron emission. Electron emitted from the
cavity wall is accelerated by RF field and attacks to the wall
again. This initial incident electron would make secondary
electron(s). The expectation number of the secondary electron(s) is
defined as yield. If the yield is more than unity and this condition is
kept during the iterations, number of electron is increased and
increased through the chain process. Finally, the cavity RF field is
totally distorted by tons of flying electrons. This process is called as
multi-pacting.

For a simple geometry like two parallel plates, the multi-pacting is
well formalized, but in general cases a simulation is only way to
examine it. In design phase, we have therefore perform a simulation to
prevent the multi-pacting in SHB.

Assumptions for the simulation are as follows; \vspace{-.2cm}
\begin{itemize}

\item RF field was evaluated by MAFIA.\vspace{-.2cm}
\item Cavity surface is copper (determins the yield).\vspace{-.2cm}
\item The initial momentum of secondary electron is 1/3 of that of the incident
      electron.\footnote{This number is experimentaly evaluated\cite{hatch}.}\vspace{-.2cm}
\item Secondary electron is emitted always perpendicular to the cavity
      surface.\vspace{-.2cm}
\item The simulation is terminated if the yield is less than unity. \vspace{-.1cm}
\item \vspace{-.2cm}If an electron is still alive after 50 processes, it is 
      defined as a possible multi-pacting condition.\vspace{-.2cm}
\end{itemize}

\begin{figure}[htbp]
\vspace{-.3cm}
\begin{minipage}{0.2\textwidth}
\includegraphics*[width=1.6in]{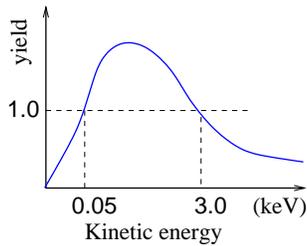}
\end{minipage}
\hfill
\begin{minipage}{0.2\textwidth}
\caption{Yield of secondary electrons as function of incident
 electron. These data were taken with copper baked by 300$^\circ
 C$\label{fig:yield}}
\end{minipage}
\vspace{-.4cm}
\end{figure}

Fig. \ref{fig:yield} shows yield of secondary electrons as function of
incident electron energy. These data are taken with copper surface baked
at 300 $^\circ C$ temperature\cite{yield}. It means that the incident
electron with its kinetic energy between 0.05 and 3.0 keV generates more
than one secondary electrons leading the multi-pacting. 

It is well known that TiN has yield less than that for copper. Because
the inner wall of the new SHB cavity is coated by TiN, then this
assumption, copper surface is conservative. Real SHB cavity is
supposedly stronger to the multi-pacting than the simulated.

With these assumptions, test particles were placed at the cavity wall
with 1.0 mm steps and cavity RF phase was scanned with 2.0 deg. step.

There are two types of multi-pacting; one point and two point. Two
point multi-pacting occurs between parallel plates, on the other hand,
one point multi-pacting occurs on a same plane. As results of the
simulation, we have found only one point multi-pacting. Basically, the
two point multi-pacting hardly occurs with external magnetic field
because the magnetic field breaks the spatial symmetry between two
plates.

The place where the one point multi-pacting was found, was the corner
edge of the cavity. In such area, the electric field is relatively weak
and electron velocity is not so fast, so that it can rotates back to the
same plane without hitting other side.

The reason that electric field becomes weak at the corner is boundary
condition on a conductor surface; electric field must be perpendicular
to the surface. Because conductor surfaces are crossing perpendicular
to each other at the corner edge, electric field perpendicular to one
surface is parallel to another surface. As the result, electric field
is going to be weak by approaching to the corner.Such area where
electric field is very weak, can be vanished by rounding the corner
edge because electric field perpendicular to the tangent of the
rounding arc is allowed.

According to this study, all of the corner of the new SHB cavity is
rounded by 5 mm radius. This 5 mm radius was determined since
multi-pacting was not found more than 3 mm away from the corner.

\section{Bunching quality}
\vspace{-.2cm}
Bunching quality which will be obtained by the new SHB cavity was
estimated by a simulation. The simulation was performed by using a
simulator for particle tracking code accounting the space charge effect, 
GPT (General Particle Tracking).

Elements set in the simulation were as follows; SHB cavities, a TW
buncher, and many Helmholz coils. The position of each element is same
as the current ATF setting, i.e. we did not perform any optimization for
the position of the elements.

The first and second SHBs are placed at 1.46 m and 2.42 m away from the
Gun exit respectively. TW buncher which has 10 cells including the
coupling cells is placed at 2.81 m. The magnetic field along the beam
axis induced by the Helmholz coils is varied from 0.005 tesla at SHB1 to
0.09 tesla at the exit of TW buncher.  Input power for SHB1 and SHB2
are set to be 6.0 and 13.0 kW respectively.

Assuming Q=7300 and R/Q=46 for the new SHB, bunch length on TW buncher
exit was obtained as 13.2 ps. This value is enough to fit S-band
acceleration.

Assuming Q=3000 and R/Q=45 corresponding to the current SHB, bunch
length on TW buncher exit was obtained as 30.0 ps. It is
larger than that required by stable S-band acceleration. It may be
a reason of the large beam loss at the injection part in ATF.

From this study, the bunching performance will be improved and the
transmission will be better by the new SHB cavity.

\section{Summary}
\vspace{-.2cm}
In ATF, the beam instability caused by injection part has been a big
issue. From a correlation study, SHB phase and amplitude fluctuations
induce the instability. The most candidate of the mechanism generating
such fluctuations is multi-pacting.

The current SHB has another problem that it has only Q value of 3000.
That is much lower than the designed value, 7800. 

Because of these reasons, we have made newly SHB cavities with careful
cosiderations; low R/Q, high Q, preventing multi-pacting.

We have manufactured SHB cavities with Q-value of 7300 which is very
close to the designed value, 7800. All of the corner of the inner wall
is rounded by 5.0 mm radius suggested by the simulation to prevent
multi-pacting.

By a particle tracking simulation, it is expected that the bunching
performance and the beam transmission of the injector system is
improved.

\end{document}